\documentclass[aps,prl,twocolumn,groupedaddress,showpacs,notitlepage]{revtex4-1}
\usepackage{graphicx}
\usepackage{amsmath,amssymb}
\usepackage{color}

\newcommand{\bra}[1]{\left<#1\right|}
\newcommand{\ket}[1]{\left|#1\right>}

\newcommand{\pd}{{\phantom{\dag}}}

\begin{document}
\title{Determining the Electron-Phonon Coupling Strength in Correlated Electron Systems from Resonant Inelastic X-ray Scattering}
\author{Luuk J. P. Ament$^{1,2}$, Michel van Veenendaal$^{3,4}$, and Jeroen van den Brink$^2$ }
\affiliation{$^1$Institute-Lorentz for Theoretical Physics, Universiteit Leiden, 2300 RA Leiden, The Netherlands}
\affiliation{$^2$Institute for Theoretical Solid State Physics, IFW Dresden, D-01171 Dresden, Germany}
\affiliation{$^3$Department of Physics, Northern Illinois University, DeKalb, IL 60115, USA}
\affiliation{$^4$Argonne National Laboratory, 9700 S Cass Ave, Argonne, IL 60439, USA}
\date{\today}


\begin{abstract}
We show that high resolution Resonant Inelastic X-ray Scattering
(RIXS) provides direct, element-specific and momentum-resolved
information on the electron-phonon (e-p) coupling strength. Our
theoretical analysis demonstrates that the e-p coupling can be extracted  from RIXS spectra by
determining the differential phonon scattering cross section. 
An alternative, very direct manner to extract the coupling is to use the one and two-phonon loss ratio, which is governed by the e-p coupling strength and the core-hole life-time. This allows measurement of the e-p coupling on an absolute energy scale.
\end{abstract}

\maketitle

Often novel electronic properties of a material can be understood by systematically unravelling the interaction between its electrons and phonons. Tunable electric transport properties in molecular crystals, for instance, are explained by the presence of a strong electron-phonon (e-p) coupling~\cite{Hulea06}. The dressing of electrons by phonons is also responsible for the colossal magnetoresistance effect in manganites~\cite{Millis95}. More delicate is the role that the e-p interaction plays in high $T_c$ superconducting cuprates -- topic of a persisting debate~\cite{Lanzara01,Giustino08,Reznik08}. The lack of a technique to measure the e-p coupling strength perpetuates this controversy. Here we show that high resolution Resonant Inelastic X-ray Scattering (RIXS) can fill this void as it gives direct, element-specific and momentum-resolved information on the coupling between electrons and phonons. We provide the theoretical framework required to distill e-p interaction strengths from RIXS, particularly in strongly correlated transition metal oxides such as the High $T_{\rm c}$ cuprates.

In RIXS experiments one scatters high energy, x-ray photons inelastically off a material~\cite{RMP}. The energy of the incident photons is chosen such that it coincides, and thus resonates, with an intrinsic electronic excitation  of the material under study -- one of the materials' x-ray absorption edges. At present the highest energy resolutions are reached at the $L$-edge of transition metal oxides, where an incident photon launches a $2p$ electron out of the atomic core into an empty $3d$ state around the Fermi-level. This highly unstable intermediate state decays rapidly, typically within 1-2 femtoseconds, so that the $2p$ core-state is refilled and an outgoing photon emitted. The state-of-the-art resolution is such that photon energy loss features on an energy scale of 25 meV can be distinguished at a copper or nickel $L_3$-edge~\cite{Ghiringhell09,Braicovich09,Braicovich10}. 

This resolution has brought phonons within the  energy window of observation and indeed last year for the first time phonon loss features were resolved in RIXS~\cite{Braicovich09, Braicovich10,Yavas09}. To put this achievement in perspective, one should realize that the incident photons at the Cu $L$-edge have an energy of around 930 eV, implying experiments have a resolving power better than $10^4$. Advanced instrumentation will drive this up further. Here we show how the progress in accuracy allows the extraction of a number of characteristics of the e-p interaction directly from RIXS, including spatial information on the e-p coupling strength. No other experimental technique has access to such e-p characteristics, particularly in strongly correlated $3d$ transition metal oxides.
The advantage of RIXS is that x-ray photons carry an appreciable
momentum. During the scattering process they can transfer it to a
phonon in the solid, so that RIXS can sample the phonon
dispersion. Present methods to measure the e-p interaction do not have
any access to ${\bf q}$-dependent information. Electron tunneling for
instance, lacks momentum-dependence and thus cannot probe the strength
of the electron-phonon coupling at a given wave vector. Moreover, it is intrinsically surface sensitive and suffers from the practical difficulty to make good yet partially transparent barriers~\cite{Allen99}. Another asset of RIXS is its element-specificity: in copper-oxides phonons can not only be accessed at the Cu $L$-edge, but also at the O $K$-edge -- resolution permitting.

Measuring phonon dispersions is an interesting new utilization of RIXS but as a technique it is up against inelastic neutron or (nonresonant) x-ray scattering. We make the case here, however, that what really sets RIXS apart is its capability to measure momentum dependent e-p couplings. To understand this capacity we start by introducing the e-p coupling Hamiltonian in generic form~\cite{Mahan00}
\begin{eqnarray}
	H^{e-p} = \sum_{{\bf k},{\bf q},\lambda} M^\pd_{{\bf q}\lambda} d^{\dagger}_{{\bf k}-{\bf q}} d^\pd_{\bf k} (b^{\dagger}_{{\bf q}\lambda}+b^\pd_{-{\bf q},\lambda}).
\label{eq:ep}
\end{eqnarray} 
The Hamiltonian couples with a strength $M_{{\bf q}\lambda}$ phonons with momentum $\bf q$ and branch index $\lambda$, which are created by the operator $b^{\dagger}_{{\bf q}\lambda}$, to electrons that are described by the fermionic operators $d^{\dagger}_{\bf k}$. The phonon dynamics are governed by the phonon Hamiltonian $H^{p} = \sum_{{\bf q},\lambda} \omega^\pd_{{\bf q}\lambda}b^{\dagger}_{{\bf q}\lambda}b^\pd_{{\bf q}\lambda}$.

For definiteness we focus on the cuprates which have all their Cu $3d$ orbitals filled apart from a single hole in the $x^2$-$y^2$ orbitals, but the analysis presented here is more general~\cite{tobepublished}.
In the Cu $L$-edge RIXS process, the $x^2$-$y^2$ state is transiently
occupied as an electron from the core is launched into it. As
mentioned above, the filled  $x^2$-$y^2$ state is very short lived and
quickly decays again. There is then a certain probability that after
the decay process a phonon is left behind in the final
state. Obviously this probability is related to the coupling of the
$x^2$-$y^2$ electron to this particular phonon. Note that the presence
of the core-hole ensures that the intermediate state is locally charge
neutral. The phonons that couple to the $3d$ $x^2$-$y^2$ quadrupolar
charge moment appear as loss-satellites in $L$-edge RIXS, due to the
absence of a charge monopole. To understand how the electron-phonon
coupling is precisely reflected in the RIXS intensity we evaluate the
RIXS amplitude for phonon scattering $A_{\bf q}$ from the Kramers-Heisenberg equation~\cite{Blume85} 
$  A_{\bf q} = 
  \sum_n \frac{\langle{f}|{\hat{D}}|{n}\rangle \langle{n}|{\hat{D}}|{i}\rangle}{\omega_{\rm{\rm det}}-E_n + i \Gamma},\label{eq:KH}
$
%
%
%
where 
$\omega_{\rm det}$ is the detuning energy of the incident photons from the $L$-edge resonance, and $E_n$ is the energy of intermediate state $n$ with respect to the resonance. The initial, intermediate and final states are $| i \rangle$, $| n \rangle$ and $| f \rangle$ and the momentum loss of the photons in the scattering process is ${\bf q}$. The dipole operators $\hat{D} \sim {\bf p}\cdot {\bf A}$ in the scattering amplitude first create a $3d$ electron out of the localized $2p$ core-state and then annihilate it. Without polarization details we can denote it as $\hat{D} =\sum_{\bf R} ( e^{-i {\bf k}_{\rm{in}}\cdot {\bf R}} d^{\dag}_{\bf R} p^{\phantom{\dag}}_{\bf R} + e^{i {\bf k}_{\rm{out}}\cdot {\bf R}} d^{\phantom{\dag}}_{\bf R}  p^{\dag}_{\bf R})$. An important quantity is the inverse core-hole life-time $\Gamma$, 
which determines the ultra-fast 1-2 fs time scale for the RIXS process.

It is well-known that due to the presence of the intermediate states,
it is in general impossible to evaluate RIXS scattering intensities
exactly, even in model systems. One therefore often resorts to
finite-size cluster calculations to compute RIXS
spectra~\cite{Kotani01,Veenendaal06}. The e-p scattering problem at hand, however, is  an exception. It can be solved exactly by means of canonical transformations. The fact that we are dealing with harmonic bosons allows for such a solution. We start by considering the limit of the coupling to an optical Einstein phonon and subsequently generalize to multiple dispersive phonons.

\begin{figure}
\begin{center}
\includegraphics[width=0.6\columnwidth]{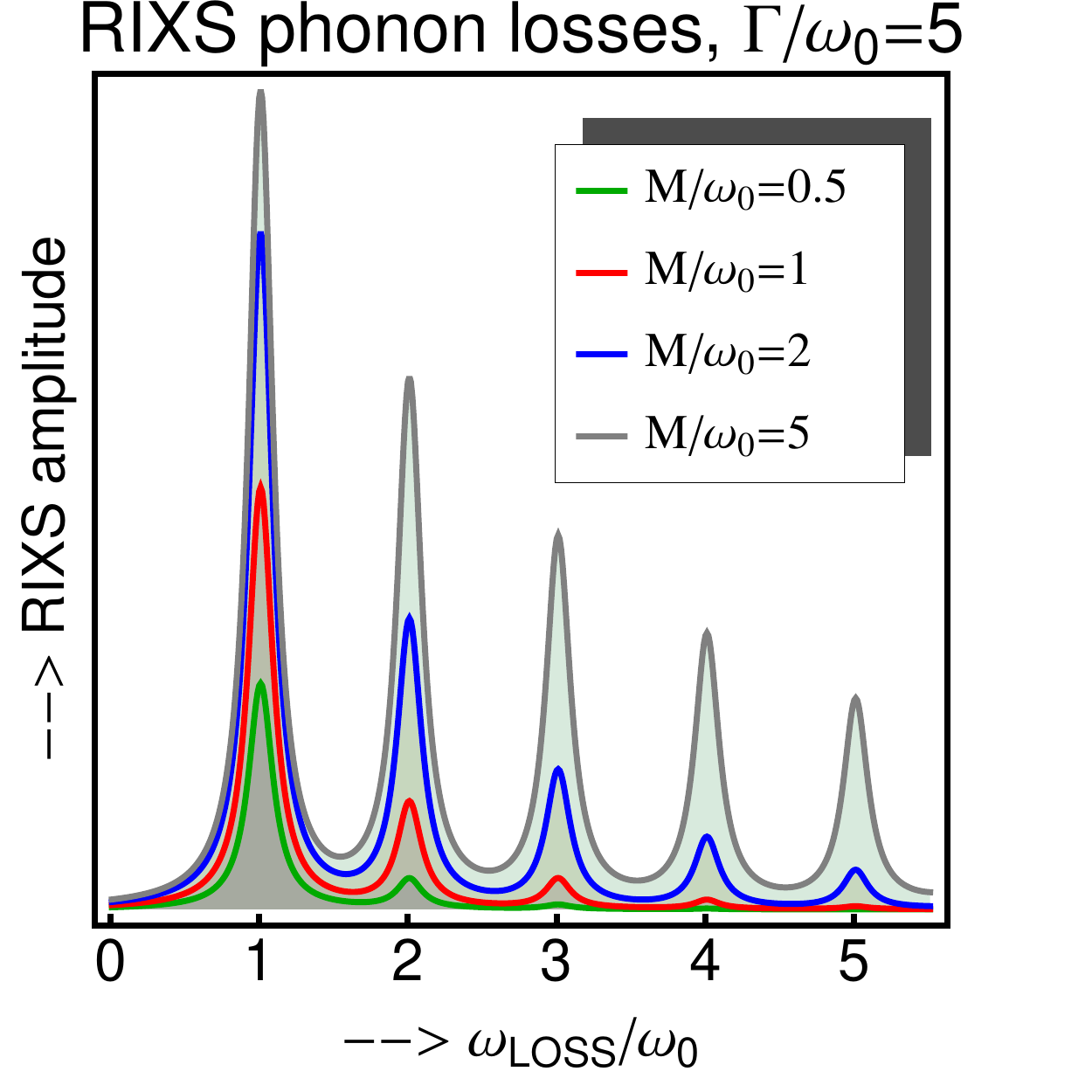}
\end{center}
\caption{Resonant Inelastic x-ray Scattering amplitude for phonon loss
  for Einstein phonon with energy $\omega_0$ for different values of
  the dimensionless electron-phonon coupling $M/\omega_0$. The inverse
  core-hole lifetime of $\Gamma/\omega_0=5$ is a typical value at the
  copper $L$-edge. The RIXS amplitude is evaluated at the incident energy that corresponds to the maximum in the x-ray absorption signal.}
\label{fig1}
\end{figure}

{\it Einstein phonon}
For a local, non-dispersive Einstein phonon the e-p Hamiltonian reduces to
$
	H = \sum_{\bf R}  M d^{\dagger}_{\bf R} d^{\phantom{\dag}}_{\bf R} (b^{\dagger}_{\bf R}+b^{\phantom{\dag}}_{\bf R}) +  \sum_{\bf R}  \omega_0 b^{\dag}_{\bf R} b^{\phantom{\dag}}_{\bf R}.
$
This Hamiltonian can be diagonalized by a canonical transformation~\cite{Mahan00} $\bar{H} = e^S H e^{-S}$, with $S=\sum_{\bf R} S_{\bf R}$ where
$
	S_{\bf R} = d^{\dag}_{\bf R} d^{\phantom{\dag}}_{\bf R}  \frac{M}{\omega_0} (b^{\dag}_{\bf R}-b^{\phantom{\dag}}_{\bf R}).
$
As there is a single core hole present, this results in
$
	\bar{H} = \sum_{\bf R} \omega_0 b^{\dag}_{\bf R} b^{\phantom{\dag}}_{\bf R} - \frac{M^2}{\omega_0},
$
where the last term merely shifts the energy at which the resonance occurs. The phonon eigenstates are all the possible occupations $\{ n_{\bf R} (m) \}$ of the phonon states with energies $E_m = \sum_{\bf R} n_{\bf R} (m) \omega_0 - M^2/\omega_0$. The eigenstates $\ket{\bar{\psi}_m}$ of $\bar{H}$ are related to the eigenstates $\ket{\psi_m}$ of $H$ by $\ket{\psi_m}= e^{-S}\ket{\bar{\psi}_m}$. Inserting this transformation in the scattering amplitude $A_{\bf q}$ gives
\begin{eqnarray}
	A^{\rm Einstein}_{\bf q} = \sum_m \frac{\bra{f}\hat{D}e^{-S}\ket{\bar{\psi}_m}\bra{\bar{\psi}_m}e^{S}\hat{D}\ket{i}}{\omega_{\rm det} - E_m + i\Gamma}  \nonumber \\
=	 \sum_{\bf R} e^{i{\bf q}\cdot {\bf R}} \sum_{n_{\bf R} = 0}^{\infty} \frac{\bra{n'_{\bf R}}e^{-S_{\bf R}}\ket{n_{\bf R}}\bra{n_{\bf R}}e^{S_{\bf R}}\ket{n^0_{\bf R}}}{z+M^2/\omega_0-n_{\bf R} \omega_0},\label{eq:Afi}
\end{eqnarray}
with $z=\omega_{\rm det} + i\Gamma$, and $n^0_{\bf R}$, $n_{\bf R}$, and $n'_{\bf R}$
are the occupations in the ground, intermediate, and final states,
respectively. The transferred momentum ${\bf q} = {\bf k}_{\rm in} -
{\bf k}_{\rm out}$ springs from the dipole operators. The
Franck-Condon  overlap factors in the numerator can be evaluated
exactly and are conveniently expressed in the functions $B_{mn}(g) =
(-1)^m \sqrt{ e^{-g} m! n!} \sum_{l=0}^n \frac{(-g)^l \sqrt{g}^{m-n}}{(n-l)! l! (m-n+l)!}$, with $g = M^2/\omega^2_0$. The RIXS amplitude to excite $n'$ phonons when starting from the ground state is
\begin{equation}
	A^{\rm Einstein}_{\bf q} = \sum_{n=0}^\infty \frac{B_{\max (n',n), \min
          (n',n)}(g)
          B_{n0}(g)}{z+(g-n)\omega_0}.
\end{equation}

\begin{figure}
\begin{center}
\includegraphics[width=0.49\columnwidth]{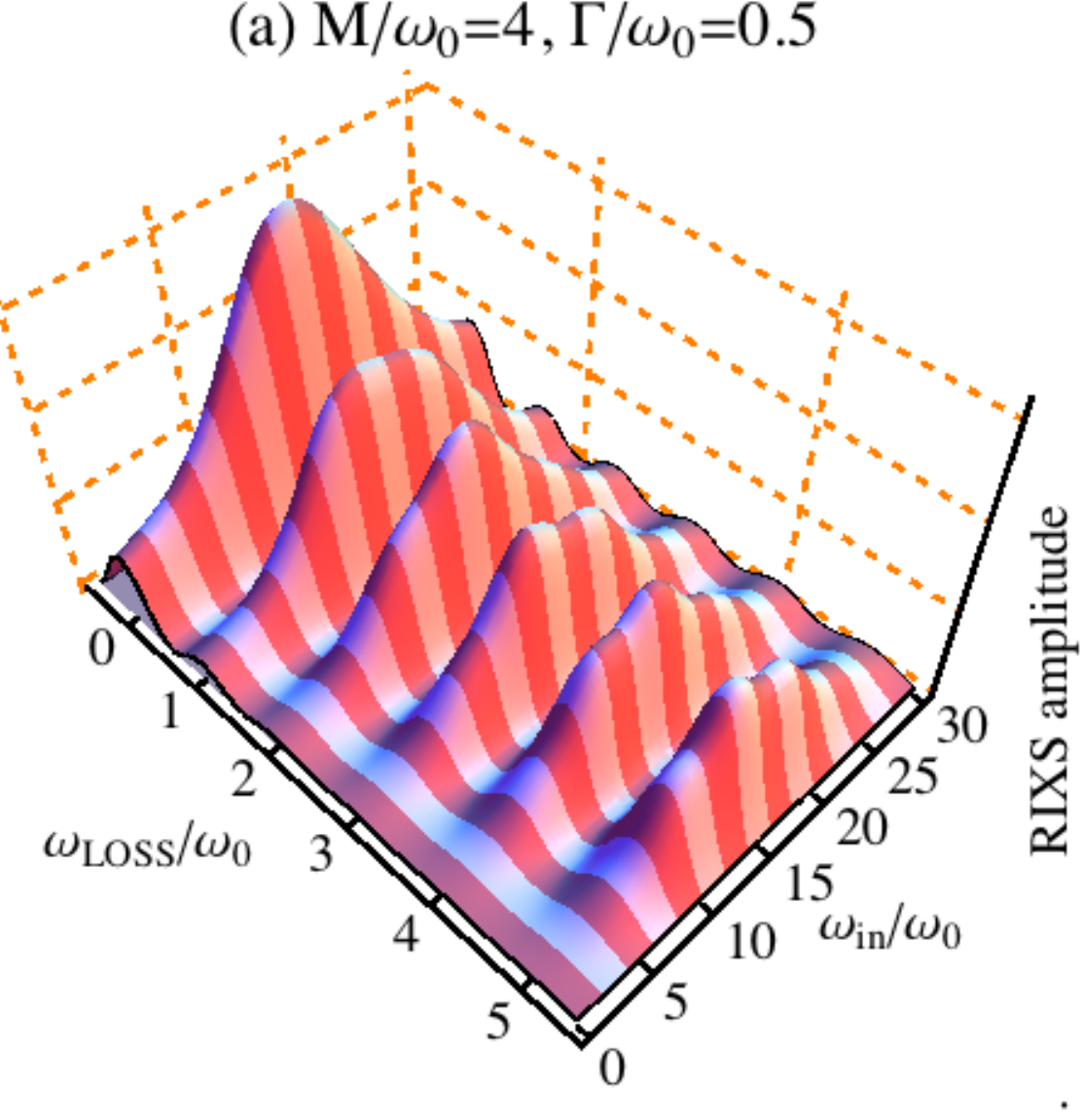}
\includegraphics[width=0.49\columnwidth]{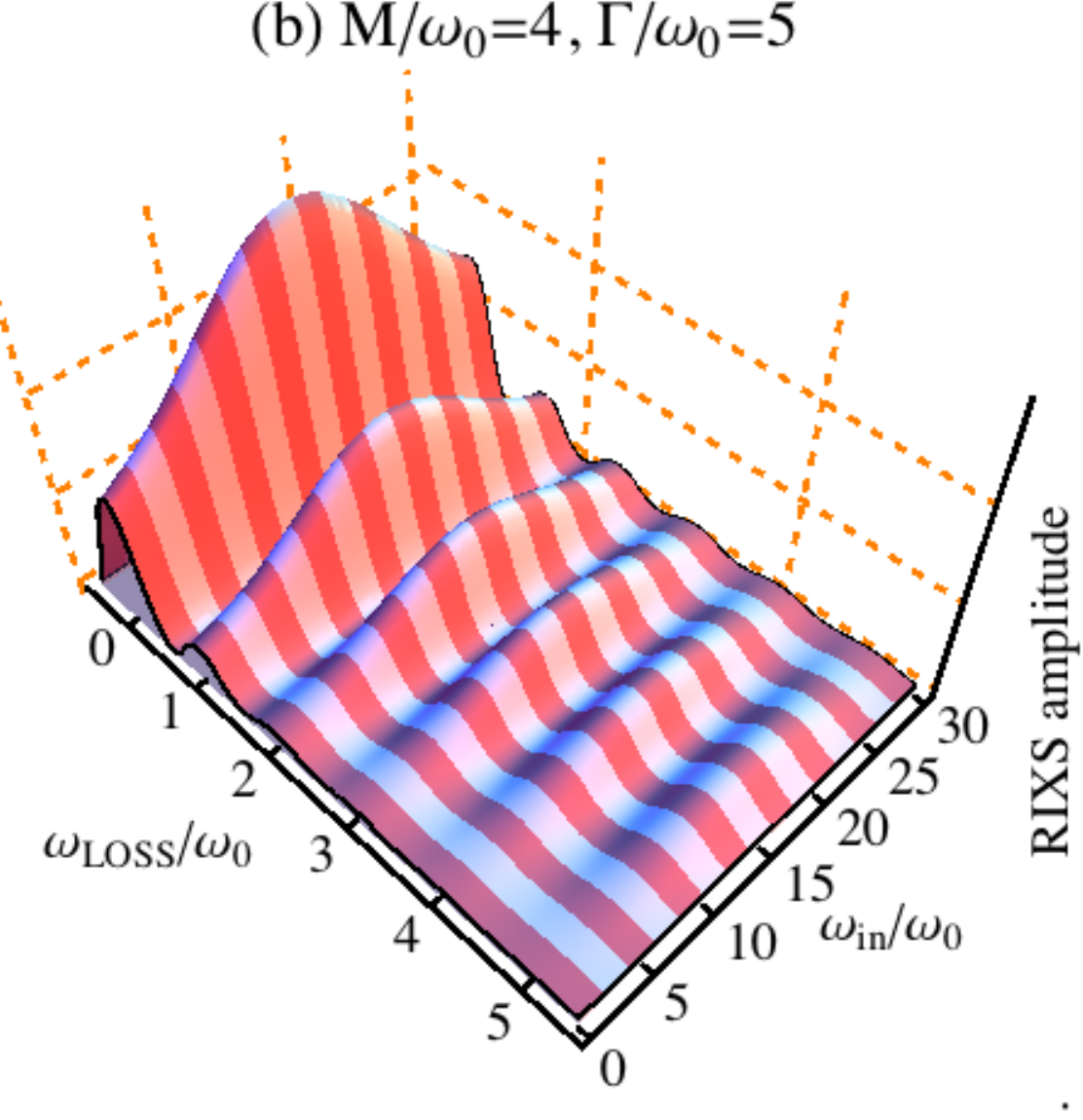}
\includegraphics[width=0.49\columnwidth]{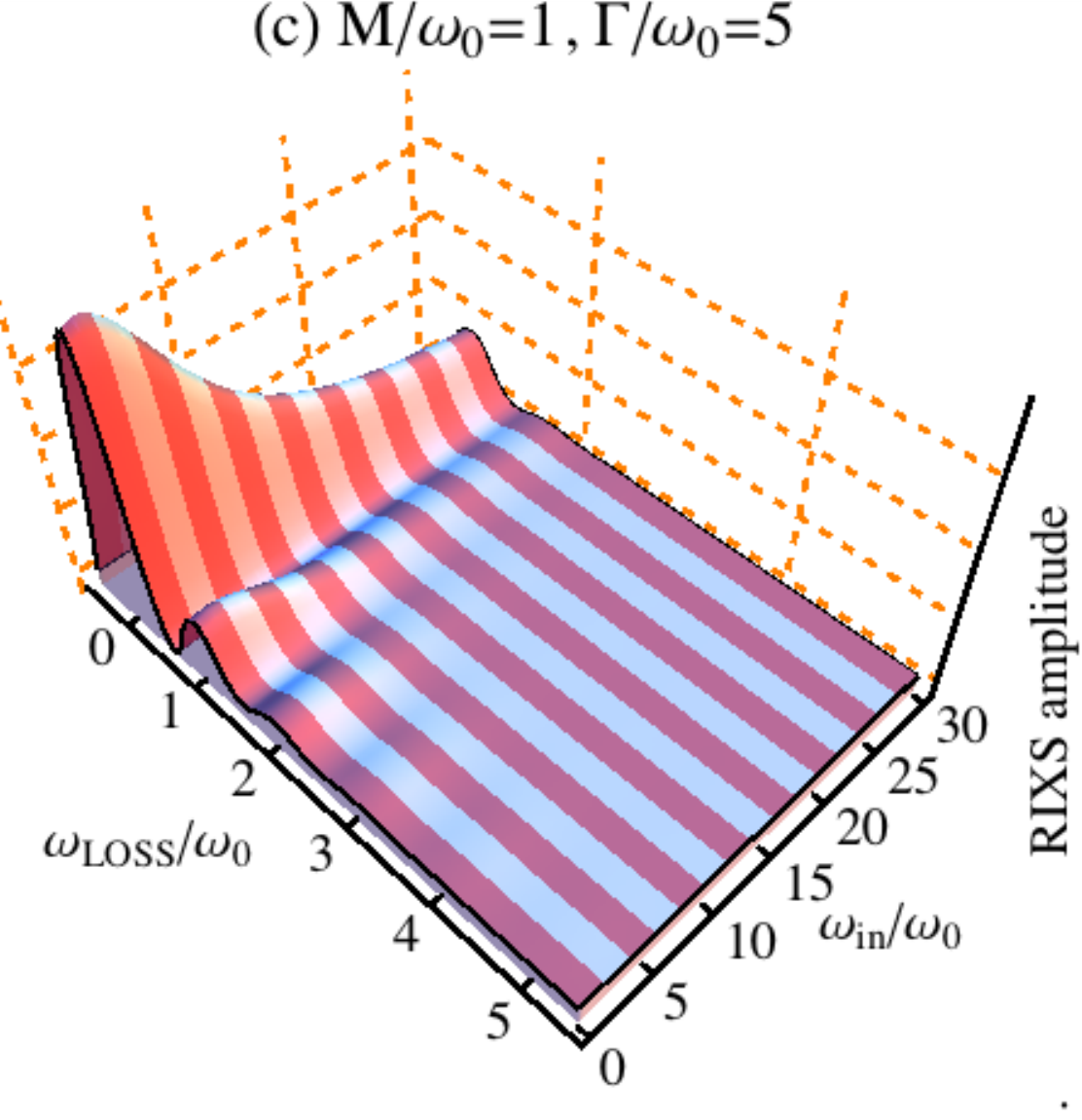}
\end{center}
\caption{Calculated RIXS amplitude for phonon loss as a function of loss energy $\omega_{loss}$ and incident energy $\omega_{in}=\omega_{det}+M^2/\omega_0$ in the case of (a) strong coupling and very long core-hole lifetime, (b) strong e-p coupling, and (c) intermediate/weak e-p coupling.}
\label{fig2}
\end{figure}

The resulting RIXS spectra for a typical weak, intermediate and strong coupling case are shown in Figs.~\ref{fig1} and~\ref{fig2}. It is clear that for stronger e-p interactions, a larger number of multi-phonon satellites carry appreciable weight. However, the first important observation is that the amplitude of the zero-loss peak $A^{(0)}$ is absolutely dominant (see Fig.~\ref{fig2}). This is even so in the strong e-p coupling regime, as long as $M/\Gamma < 1$. As at the Cu $L_3$-edge $\Gamma=280$ meV~\cite{Krause79}, this corresponds to the physical situation. This observation is empirically supported by the fact that a dispersion of magnon~\cite{Braicovich09} and bimagnon~\cite{Hill08,Brink07,Forte08a} excitations have been observed in $L$- and $K$-edge RIXS. Such dispersion cannot be present if these magnetic excitations are always accompanied by (multiple) phonon excitations that carry away momentum.

\begin{figure}
\begin{center}
\includegraphics[width=0.6\columnwidth]{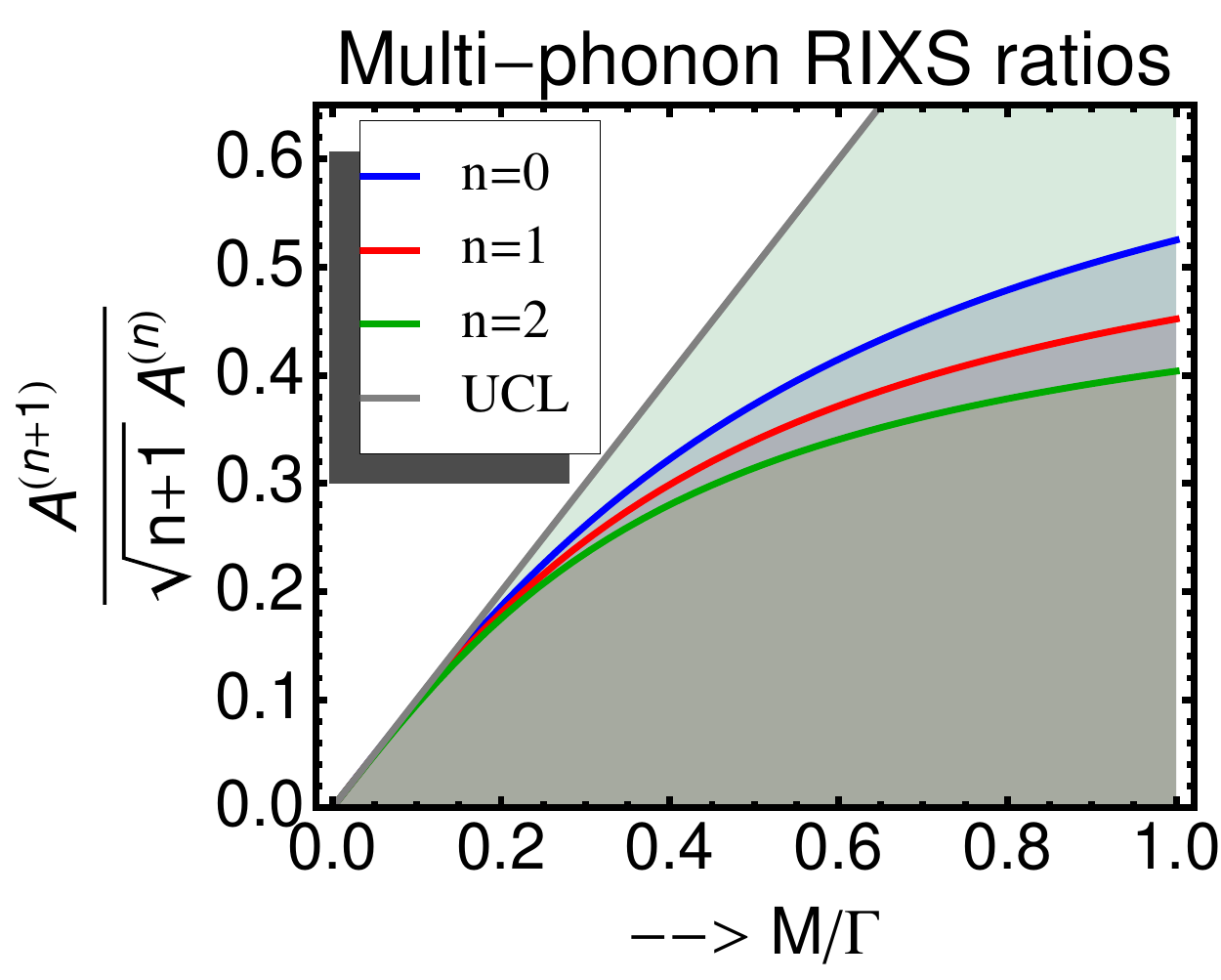}
\end{center}
\caption{Relation of the multiple phonon resonant RIXS amplitude to the electron-phonon coupling strength $M/\Gamma$. The ratio of the $n+1$ phonon $A^{(n+1)}$ to $n$ phonon $A^{(n)}$ loss amplitudes is shown. $\Gamma$ is the inverse core-hole life-time. In the physical relevant regime these curves do not depend on the phonon frequency $\omega_0$. The UCL expansion (straight gray line) gives accurate results for $M/\Gamma \lesssim 0.2$. 
}
\label{fig3}
\end{figure}

The exact amplitude for exciting a single phonon in the RIXS
process is $A^{(1)}= (e^{-g}/ \sqrt{g} )\sum^{\infty}_{n=0} g^n
(n-g)/[n! (z+ (g-n) \omega_0 )]$. In leading order in the e-p coupling constant this amplitude is 
%
%
$A^{(1)}=M / z^2$. The differential single-phonon RIXS scattering amplitude $A^{(1)}$ is therefore directly proportional to the e-p coupling constant $M$, see Fig.~\ref{fig3}. By increasing $z$ and thus tuning away from the absorption edge the scattering amplitude is reduced. 

{\it Dispersive phonons.}
In close analogy with Eq.~(\ref{eq:Afi}), the case of multiple dispersive phonon modes can also be solved exactly. The scattering amplitude is
\begin{eqnarray}
 &&A_{\bf q} =  \sum_{\bf R} e^{i{\bf q}\cdot {\bf R}} \sum_m  \bigg[  \nonumber \\
  &&  \left. \frac{\prod_{{\bf k},\lambda} \bra{n'_{{\bf k}\lambda}}e^{-S_{{\bf R}{\bf k}\lambda}}\ket{n_{{\bf k}\lambda}(m)}\bra{n_{{\bf k}\lambda}(m)}e^{S_{{\bf R}{\bf k}\lambda}}\ket{n^0_{{\bf k}\lambda}}}{z+\sum_{{\bf k},\lambda}[g_{{\bf k}\lambda}- n_{{\bf k}\lambda}(m)] \omega_{{\bf k}\lambda}} \right]
  \label{eq:KH}
\end{eqnarray}
where $n^0_{{\bf k}\lambda}$, $n_{{\bf k}\lambda}$, and $n'_{{\bf  k}\lambda}$ are the occupation numbers of the modes indexed by
${\bf k}$ and $\lambda$ in the ground, intermediate, and final states,
respectively. The sum over $m$ is over all intermediate state
occupations. Together with the Franck-Condon factors this constitutes an exact, closed expression for the RIXS response.

{\it UCL approximation}
Even when the e-p coupling $g$ is not small, one can obtain the RIXS amplitude in approximate form by using the fact that the time-scale of a typical phonon ($80$ meV $\sim 52$ fs) is much slower than the ultrashort RIXS time-scale ($1.6$ eV $\sim 2.6$ fs at the Cu $K$-edge). This separation of time-scales suggests that the scattering process contains a viable expansion parameter, which is small even for a fast phonon. The Ultra-short Core-hole Lifetime (UCL) expansion formalizes this observation~\cite{Brink06,Ament07}. In the present case it boils down to an expansion of the Kramers-Heisenberg expression in terms of $M_{{\bf q}\lambda}/\Gamma$.
With the UCL expansion we obtain a compact, approximate expression for the phonon scattering amplitude that is also valid at finite temperature. In the case of dispersive phonons, Eq.~(\ref{eq:KH}) is expanded as
\begin{align}
&A_{{\bf q},UCL} \approx \frac{1}{z} \sum_{\bf R} e^{i{\bf q}\cdot {\bf R}} \prod_{{\bf k},\lambda} \sum_{n_{{\bf k} \lambda}=0}^{\infty} \bra{n'_{{\bf k}\lambda}}e^{-S_{{\bf R}{\bf k}\lambda}}\ket{n_{{\bf k}\lambda}} \nonumber \\
	&~~~\times \bra{n_{{\bf k}\lambda}}e^{S_{{\bf R}{\bf k}\lambda}}\ket{n^0_{{\bf k}\lambda}} \sum_{l=0}^{\infty} \left( (n_{{\bf k}\lambda}-g_{{\bf k}\lambda}) \omega_{{\bf k}\lambda}/z \right)^l
\end{align}
where $S_{{\bf R}{\bf k}\lambda} = d^\dag_{\bf R} d^\pd_{\bf R} \tfrac{M_{{\bf k}\lambda}}{\omega_{{\bf k}\lambda}} e^{i{\bf k}\cdot {\bf R}} (b^\dag_{-{\bf k}\lambda} - b^\pd_{{\bf k}\lambda})$ and $g_{{\bf k}\lambda} = |M_{{\bf k}\lambda}/\omega_{{\bf k}\lambda}|^2$. 
As $z$ is large compared to the phonon and e-p energy scales, we retain the leading terms in $l$ and find the inelastic RIXS amplitude
\begin{equation}
	A^{(1)}_{{\bf q},{UCL}} = \frac{M_{{\bf q}\lambda}}{z^2} 
	\bra{f} b^{\dag}_{{\bf q}\lambda} + b^{\phantom{\dag}}_{-{\bf q},\lambda} \ket{i}.
\end{equation}
to first order in $M_{{\bf k}\lambda}/z$. Again, multi-phonon contributions to $A_{\bf q}$ are suppressed proportional to $ 
M_{{\bf k}\lambda}/\Gamma$. 
%
%
%
This result implies that {\it momentum dependent RIXS can directly map out the $\bf q$-dependence of the e-p coupling strength  $M_{{\bf q}\lambda}$}. From this information for instance the spatial range of the e-p interaction $M_{{\bf r}{\lambda}}$ can be determined as it is directly related to the Fourier transform of $M_{{\bf q}{\lambda}}$.

%

RIXS provides still another method to extract e-p coupling strengths, which is expected to be particularly  powerful in the case of weakly dispersive optical phonons, for instance modes around 80 meV in the high $T_{\rm c}$ cuprates~\cite{Astuto02,Padilla05}. From the UCL expansion one finds that the ratio of the one and two-phonon loss amplitude also directly reflects the e-p coupling constant:  $A^{(2)}/A^{(1)} = \sqrt{2} M/z$. This is confirmed by the exact solution, for $M/\Gamma \ll 1$, see Fig.~\ref{fig3}. For strongly dispersive phonons this method also works if both $A^{(1)}_{\bf q}$ and $A^{(2)}_{\bf q}$ are measured throughout the Brillouin zone. 


In the analysis above we concentrated on transition metal $L$-edge RIXS, which has the advantage of a photo-electron launched directly into the $3d$ state. A certain $3d$ orbital can be selected by choosing the polarization of incident and outgoing x-rays~\cite{Kuiper98,Ament09}, so that e-p characteristics related this particular $3d$ orbital can be measured~\cite{Braicovich09}. It is straightforward to include $d$-$d$ excitations that couple to breathing and Jahn-Teller phonons~\cite{Brink01}, which is of great importance in the study of Jahn-Teller polarons. Also  at the O $K$-edge phonons can be probed but as this edge is at lower energy, the photons have less momentum and a smaller part of the Brillouin zone can be probed, which also holds for Cu M-edges. As hard x-ray transition metal $K$-edges do not suffer this disadvantage they do provide in principle a viable method to measure extensively momentum dependent phonon properties and e-p interactions~\cite{Yavas09,Hancock10}.

The framework presented here to extract the e-p coupling interaction bestows on RIXS the unique potential to provide direct, element-specific and momentum-resolved information on the interaction between electrons and phonons {\it on an absolute scale}. In weakly correlated electron systems these properties can be computed with modern {\it ab initio} electronic structure methods, for instance in the newly discovered iron pnictide superconductors~\cite{Boeri08}, and our framework to distill them from RIXS allows a direct comparison. In strongly correlated materials, particularly the high $T_{\rm c}$ cuprates, these assets make high resolution RIXS  a unique tool to unravel the interaction between its electrons and phonons. 

We thank Lucio Braicovich, John Hill, Steven Johnston and Tom Devereaux for fruitful discussions. This work is supported by the U.S. Department of Energy, Office of Basic Energy Sciences under contract DE-AC02-76SF00515 and benefited from the RIXS collaboration supported by the Computational Materials Science Network (CMSN) program under grant number DE-FG02-08ER46540. This work is supported by the Dutch "Stichting voor Fundamenteel Onderzoek der Materie" (FOM).
MvV was supported by the U.S. Department
of Energy (DOE), No. DE-FG02-03ER46097. Work at Argonne National
Laboratory was supported by the U.S. DOE, Office of Basic Energy Sciences (BES), 
under contract No. DE-AC02-06CH11357.
This research benefited from the RIXS collaboration supported by the
Computational Materials Science Network (CMSN), BES, 
DOE under grant number DE-FG02-08ER46540.



\bibliographystyle{prsty}

\begin{thebibliography}{10}

\bibitem{Hulea06}
I.~N. Hulea {\it et~al.}, Nature Materials {\bf 5},  982  (2006).

\bibitem{Millis95}
A.~J. Millis, P.~B. Littlewood, and B.~I. Shraiman, Phys. Rev. Lett. {\bf 74},
  5144  (1995).

\bibitem{Lanzara01}
A. Lanzara {\it et~al.}, Nature {\bf 412},  510  (2001).

\bibitem{Giustino08}
F. Giustino, M.~L. Cohen, and S.~G. Louie, Nature {\bf 452},  975  (2008).

\bibitem{Reznik08}
D. Reznik, G. Sangiovanni, O. Gunnarsson, and T.~P. Devereaux, Nature {\bf
  455},  E6  (2008).

\bibitem{RMP}
L. Ament {\it et~al.}, ArXiv:1009.3630  (2010).

\bibitem{Ghiringhell09}
G. Ghiringhelli {\it et~al.}, Phys. Rev. Lett. {\bf 102},  027401  (2009).

\bibitem{Braicovich09}
L. Braicovich {\it et~al.}, Phys. Rev. Lett. {\bf 102},  167401  (2009).

\bibitem{Braicovich10}
L. Braicovich {\it et~al.}, Phys. Rev. Lett. {\bf 104},  077002  (2010).

\bibitem{Yavas09}
H. Yavas {\it et~al.}, ArXiv:1009.4354  (2010).

\bibitem{Allen99}
P. Allen, {\em Handbook of Superconductivity} (Academic Press, New York,
  ADDRESS, 1999), p.\ 478.

\bibitem{Mahan00}
G.~D. Mahan, {\em Many-Particle Physics} (Springer-Verlag, New York, ADDRESS,
  2000).

\bibitem{tobepublished}
L. Ament, M. van Veenendaal, and J. van~den Brink, to be published  .

\bibitem{Blume85}
M. Blume, J. Appl. Phys. {\bf 57},  3615  (1985).

\bibitem{Kotani01}
A. Kotani and S. Shin, Rev. Mod. Phys. {\bf 73},  203  (2001).

\bibitem{Veenendaal06}
M. van Veenendaal, Phys. Rev. Lett. {\bf 96},  117404  (2006).

\bibitem{Krause79}
M.~O. Krause and J.~H. Oliver, J. Phys. Chem. Ref. Data {\bf 8},  329  (1979).

\bibitem{Hill08}
J.~P. Hill {\it et~al.}, Phys. Rev. Lett. {\bf 100},  097001  (2008).

\bibitem{Brink07}
J. van~den Brink, Europhys. Lett. {\bf 80},  47003  (2007).

\bibitem{Forte08a}
F. Forte, L.~J.~P. Ament, and J. van~den Brink, Phys. Rev. B {\bf 77},  134428
  (2008).

\bibitem{Brink06}
J. van~den Brink and M. van Veenendaal, Europhys. Lett. {\bf 73},  121  (2006).

\bibitem{Ament07}
L.~J.~P. Ament, F. Forte, and J. van~den Brink, Phys. Rev. B {\bf 75},  115118
  (2007).

\bibitem{Astuto02}
M. d'Astuto {\it et~al.}, Phys. Rev. Lett. {\bf 88},  167002  (2002).

\bibitem{Padilla05}
W.~J. Padilla, M. Dumm, and D.~N. Basov, Phys. Rev. B {\bf 72},  205101
  (2002).

\bibitem{Kuiper98}
P. Kuiper {\it et~al.}, Phys. Rev. Lett. {\bf 80},  5204  (1998).

\bibitem{Ament09}
L.~J.~P. Ament {\it et~al.}, Phys. Rev. Lett. {\bf 103},  117003  (2009).

\bibitem{Brink01}
J. van~den Brink, Phys. Rev. Lett. {\bf 87},  217202  (2001).

\bibitem{Hancock10}
J.~N. Hancock, G. Chabot-Couture, and M. Greven, New Journal of Physics {\bf
  12},  033001  (2010).

\bibitem{Boeri08}
L. Boeri, O.~V. Dolgov, and A.~A. Golubov, Phys. Rev. Lett. {\bf 101},  026403
  (2008).

\end{thebibliography}

\end{document}